# Ampliando horizontes sobre medición del desempeño y el concepto de regularidad en el baloncesto profesional

Salmerón-Gómez, Román. Departamento de Métodos Cuantitativos. Universidad de Granada. romansg@ugr.es

Gómez-Haro, Samuel. Departamento de Organización de Empresas. Universidad de Granada. samugh@ugr.es

**Resumen**

Conseguir información relevante sobre el desempeño de equipos y sus jugadores para la toma de decisiones lleva provocando un continuo análisis de aquellos elementos relevantes que afectan al resultado individual y del conjunto en al ámbito del deporte. La obligación de obtener resultados positivos hace necesaria usar herramientas de gestión que ayuden a la estimación fiable del rendimiento deportivo de los jugadores profesionales de cara a la toma de decisiones por los dirigentes, por ejemplo, en el momento de diseñar una plantilla. El presente trabajo trata de mejorar el análisis existente presentando una metodología para medir el rendimiento y regularidad de los jugadores profesionales de baloncesto. Utilizamos una muestra de datos de los 15 mejores jugadores en términos de desempeño según la primera liga española de baloncesto ACB en la pasada temporada 2013/2014 para redefinir y mostrar nuevos elementos a tener en cuenta en la valoración del desempeño de los jugadores. Los resultados obtenidos presentan perspectivas interesantes para aspectos de la gestión deportiva como el reclutamiento, diseño de equipos y desempeño de las organizaciones.

## 1. Introducción

Uno de los objetivos dentro del amplio campo de la gestión deportiva ha sido descubrir herramientas que provean a los entrenadores y directivos de organizaciones de un conocimiento más profundo de los recursos que dispone de cara a mejorar el rendimiento de los mismos. Esas funciones incluyen el diseño y la construcción de las plantillas, organización de los programas de entrenamiento, planificación de estrategias de temporada y de juego; y la relación de los jugadores-equipo con su desempeño. La información previa nos sirve para conocer las fortalezas y debilidades a nivel individual y de equipo, propio y del oponente con el que te enfrentas cada jornada, y el desempeño en las diferentes situaciones existentes en un determinado contexto complejo como son los equipos deportivos profesionales. El análisis del desempeño es un tema que siempre ha estado en continuo desarrollo, pero desde el caso de Moneyball (Lewis, 2004) vivimos en una continua revisión de los conceptos existentes y búsqueda de mejores técnicas de análisis de la información que ofrezca información valiosa para la toma de decisiones. En este trabajo desarrollamos nuevos instrumentos fiables que nos provean de una nueva visión del desempeño individual de los jugadores en el juego colectivo dentro del contexto del deporte del baloncesto profesional, utilizando como muestra la primera división española de baloncesto profesional (Liga ACB).

El baloncesto, como deporte de equipo que es, se caracteriza por la ejecución de una serie de habilidades en múltiples situaciones que surgen durante el desarrollo del juego. Con cada habilidad, existen infinito número de posibles soluciones que pueden ser tomadas, que dependen de las características de los jugadores, de su previa experiencia en el juego, etc. La mayoría de esas acciones tienen un determinado carácter (acciones ofensivas o defensivas por ejemplo), así que las posibilidades de tener éxito son variadas. El diseño de las plantillas deportivas es un ejemplo perfecto de sistemas dinámicos que sirven para explicar los mecanismos de autoregulación de los jugadores, los cuales varían y ajustan su comportamiento acorde a lo que suceden durante el desarrollo del juego (Gutiérrez y Ruiz, 2013). Además, los deportes colectivos se ajustan dinámicamente acorde a las características de los equipos que disputan el partido y de los jugadores propios y del rival (Tminic, Tminic & Papic, 2009). Para optimizar el resultado de un equipo es necesario considerar el grado de dominio del juego a nivel colectivo, el desempeño de cada jugador, las relaciones con compañeros y oponentes. Así que tenemos un sistema complejo de variables que son imposibles de controlar en su totalidad, donde la creación de mecanismos autorregulatorios de estos sistemas dinámicos son los adecuados para una mejor comprensión.

Por tanto, en el proceso de preparación de un equipo, el entrenador debe facilitar las condiciones para el desarrollo de las habilidades de los jugadores y el modelo táctico preciso que permita obtener la mejor eficacia y eficiencia posible en el juego (Tmini et al., 2009). Así, se convierte en necesario saber los apropiados niveles de desempeño de cada jugador acorde a sus propias características, su nivel de habilidades y la posición en la que juegan para guiar su preparación y que puedan explotar sus fortalezas y mejorar sus debilidades.

El trabajo se estructura como sigue: en la sección 2 se establece la metodología de trabajo para medir el desempeño del jugador. En la sección 3 se proponen nuevas formas de medir el rendimiento de los jugadores, utilizando como muestra los quince jugadores mejor valorados por la liga ACB en la temporada 2013/2014, mientras que en la sección 4 se propone usar la dispersión de los datos para medir la regularidad de los mismos. En la sección 5 se proponen técnicas de inferencia estadística para medir la importancia de un jugador dentro de un equipo. En la sección 6 se muestran los resultados obtenidos considerando los 15 mejor valorados de la Liga ACB en la temporada 2013/14. Finalmente, en la sección 7 se establecen las principales conclusiones del presente trabajo.

## 2. Medición del desempeño

La primera liga española ACB de baloncesto es considerada como unas de las mejores competiciones en este deporte, donde la posibilidad de atraer jugadores con talento es alta y posee una cobertura de medios atractiva. En esta situación, todos los equipos comparten el mismo objetivo inicial, ganar el máximo número de partidos, y diseñan sus plantillas y estrategias en base a ese objetivo.

En el momento de diseñar una plantilla de cualquier equipo deportivo se tienen en cuenta diversos factores, como por ejemplo, filosofía de juego, compatibilidad de jugadores y, cómo no, recursos económicos. Puesto que estos últimos son limitados es importante disponer de la mayor información posible para tomar la decisión correcta.

El juego del baloncesto genera muchísima información numérica, de ahí el gran y diverso número de índices de valoración de jugadores. Así, por ejemplo, en Martínez (2010a) se recogen 229 sistemas para medir el rendimiento de un jugador clasificados fundamentalmente según: a) si miden producción defensiva, ofensiva o ambas y b) si son obtenidos a partir del boxscore o del play by play; lo que varía el nivel de complejidad para su cálculo. Usar únicamente uno de ellos nos provee una visión muy limitado del desempeño siendo mejor opción usar un número reducido[1] de los mismos.

---

[1] Tan nocivo puede resultar ignorar toda información estadística como pretender usarla toda. Buscar un equilibrio puede ser la clave del éxito. Es más, dicho equilibrio no tiene por que ser universal, en potencia existe una distinta combinación de índices para cada decisor: aquellos que se adecúen mejor a su filosofía de ver el juego.

Por tal motivo, el uso de un único índice en competiciones como la Liga ACB o Euroliga no parece lo más adecuado.

Aunque a continuación se propongan índices para medir el rendimiento de un jugador, pretender establecer un ranking sobre los índices que se deben usar es una cuestión que se encuentra totalmente fuera del objetivo de este trabajo. Sin embargo, si deseamos establecer una metodología de trabajo para medir el desempeño:

1. Seleccionar los índices que se consideren más oportunos para medir el desempeño. En este caso se propone un nuevo índice.
2. Usar los valores medios del índice anterior y el concepto de dispersión para analizar y medir el concepto de regularidad del jugador.
3. Usar técnicas de inferencia estadística para analizar la influencia de un jugador en el rendimiento del equipo.

Destacar que los índices propuestos en el presente trabajo están basados en la información disponible en el boxscore. Si bien somos conscientes de que esta cuestión presenta muchas limitaciones que pueden ser mejoradas o resueltas mediante el play by play, el hecho de que esta última información no esté disponible en muchos países ni en todas las categorías nos hace decantarnos por la primera opción de cara a conseguir una utilidad máxima del mismo. La publicación de dichos índices en el trabajo pretende exponer y compartir reflexiones y aportaciones al campo del desempeño en equipos profesionales de baloncesto para su amplio uso.

Finalmente, para centrar el estudio se compararán los índices propuestos con la valoración usada en la Liga ACB y Euroliga.

## 3. Análisis del rendimiento de un jugador

En la presente sección se propone un nuevo índice para medir el rendimiento de un jugador. Además, se pone de manifiesto la importancia de promediar los valores medios de dicho índice (y cualquier otra característica: puntos, rebotes, etc) en función de los minutos disputados por cada jugador.

Es práctica profesional habitual que el uso de los datos de rendimiento se utilice, además de para medir el desempeño de un jugador a lo largo de un partido, como una referencia más a la hora de renovar o fichar jugadores. Ahora bien, dado su uso, para que sea realmente útil y conduzca a decisiones correctas es imprescindible que dicha información sea tratada correctamente. En este sentido la valoración usada en la Liga ACB o Euroliga presentan limitaciones tales como redundancia y que sólo son obtenidas a partir del boxscore (ver Martínez, 2010c). Además, no se tiene en cuenta un factor tan importante en la acumulación de acciones como es el tiempo jugado.

**3.1 Propuesta para medir el rendimiento de un jugador**

La creación de índices de rendimiento es una cuestión muy sensible y que de cara a su comprensión y mayor utilidad consideramos que se debe basar en dos principios concretos:

- Debe de poder calcularse fácilmente: si bien somos conscientes que limitarse a calcular un índice a partir de una combinación lineal de la información proporcionada por el boxscore supone no recoger todo lo que sucede en la cancha, la realidad es que la información obtenida a partir del play by play sólo es recogida en algunos países y no en todas sus competiciones. Por ejemplo, en España si está disponible la información del play by play, pero sólo para su máxima competición[2], dejando de lado el resto de categorías existentes. El índice propuesto debe de poder ser computable en cualquier liga, independientemente de su nivel de profesionalización y, por tanto, de recursos.
- Debe ser capaz de medir distintos aspectos del juego: como se ha comentado limitarse a usar un único índice no parece que sea la mejor idea. Por tal motivo, el índice propuesto será el resultado de la agregación de otros dos, uno que mida el rendimiento defensivo del jugador y otro el ofensivo.

Más concretamente, el índice propuesto es:

Rendimiento (Rend) = Índice Defensivo (ID) + Índice Ofensivo (IO),

donde:

ID = Rebotes defensivos (RD) + Tapones a favor (TF) - Faltas personales cometidas (FPC) + 2 * Balones recuperados (BR),

IO = Tiros de 2 convertidos (T2C) + Tiros de 1 convertidos (T1C) + 1.5 * Tiros de 3 convertidos (T3C) - Tiros de 2 fallados (T2F) - 2 * Tiros de 1 fallados (T1F) - Tiros de 3 fallados (T3F) + 2 * Rebotes ofensivos (RO)

---

[2] Según Martínez (2010b) en la competición LEB Oro (actual ADECCO Oro) el play by play sí está disponible para los equipos. Aunque nosotros no tenemos constancia que lo esté para el público en general.

+ 2 * Asistencias (A) + 1.5 * Faltas personales recibidas (FPR) - 2 * Balones perdidos (BP).

La elección de las ponderaciones no responde a ningún criterio estadístico/econométrico, lo cual podría ser considerado como un inconveniente. Sin embargo, a nuestro parecer se trata de una ventaja ya que al ser subjetivos son totalmente interpretables:

- Las acciones positivas suman y las negativas restan.
- Las ponderaciones por defecto son 1 excepto en aquellos casos que se desee premiar o penalizar con más intensidad cierta acción.
- El coeficiente asociado a BR es 2 ya que se recupera la posesión y se evita que el equipo rival lance a canasta.
- El coeficiente asociado a T3C es de 1.5 debido a la mayor dificultad del tiro de larga distancia y mayor rédito cuando se convierte.
- El coeficiente asociado a T1F es de -2 con el objetivo de penalizar con mayor peso una acción de gran importancia en el baloncesto y que siempre se ejecuta en las mismas circunstancias.
- El coeficiente asociado a RO es 2 ya que se evita una posesión del rival y se consigue una opción más de tiro.
- El coeficiente asociado a A es de 2 con el objetivo de premiar doblemente una acción del juego en la que se consigue una canasta fácil.
- El coeficiente asociado a FPR es de 1.5 ya que (como mínimo) carga al jugador rival con una falta y acerca al equipo al bonus (es decir, que toda falta personal suponga tiros libres).
- El coeficiente asociado a BP es -2 ya que se trata de una posesión donde no se ha podido lanzar y supone una posesión extra para el rival.

En la Tabla 1 se recogen los valores obtenidos para el cálculo del *Rendimiento* en los primeros 15 jugadores del ranking ACB en cuanto valoración se refiere de la liga regular 2013/14 (incluidos playoff por el título). Llama la atención la subida experimentada en el ranking por Rodríguez y Satoransky (con diferencia los dos mejores asistentes en el grupo y de los que más balones recuperan) y la bajada de Panko (el que más tiros de campo falla). Este segundo caso cobra mayor importancia si se tiene en cuenta que Panko es el máximo anotador de los 15. Su mejor clasificación en la Valoración ACB se debe especialmente a los puntos anotados aunque estos se hayan producido con errores de tiro.

Tabla 1. Valoración ACB y rendimiento medios de los primeros 15 clasificados temporada 2013/14

| Jugador | Equipo | Valoración media | Posición | Rendimiento medio | Posición |
|---|---|---|---|---|---|
| **Rodríguez** | Real Madrid | 14.55 | 11 | 15.58 | 5 |
| **Doblas** | Gipuzkoa Basket | 13.85 | 15 | 13.53 | 10 |
| **Doelman** | Valencia Basket | 17.33 | 2 | 16.76 | 2 |
| **Gabriel** | Bilbao Basket | 14.76 | 9 | 13.19 | 11 |
| **Kirksay** | Joventut Badalona | 14.42 | 12 | 14.05 | 8 |
| **Llull** | Real Madrid | 15.2 | 7 | 14.43 | 7 |
| **Mirotic** | Real Madrid | 14.7 | 10 | 13.83 | 9 |
| **Mumbrú** | Bilbao Basket | 14.03 | 14 | 11.96 | 14 |
| **Nocioni** | Laboral Kutxa | 14.33 | 13 | 10.05 | 15 |
| **Panko** | Fuenlabrada | 16.91 | 3 | 12.8 | 13 |
| **Pleiss** | Laboral Kutxa | 17.47 | 1 | 15.92 | 8 |
| **Robinson** | Gipuzkoa Basket | 16.21 | 4 | 12.94 | 12 |
| **Rudy** | Real Madrid | 15.64 | 6 | 15.96 | 3 |
| **Satoransky** | Cajasol | 15.81 | 5 | 17.43 | 1 |
| **Tomic** | FC | 15.07 | 8 | 14.51 | 6 |

| | Barcelona | | | | |
|---|---|---|---|---|---|

En relación a éste último comentario, destacar que la correlación existente entre los puntos anotados y la Valoración ACB es mayor que entre los puntos anotados y *Rendimiento* (ver Tabla 2) en las tres medidas de correlación usadas[3]. En todos los casos las correlaciones son significativamente[4] distintas de cero.

Tabla 2. Correlaciones entre Valoración ACB y Rendimiento con los puntos anotados

| Correlaciones | Pearson | Kendall | Spearman |
|---|---|---|---|
| **Valoración ACB con puntos anotados** | 0.752 | 0.514 | 0.697 |
| **Rendimiento con puntos anotados** | 0.504 | 0.294 | 0.421 |

Este aspecto pone de relevancia que la Valoración ACB beneficia a aquellos jugadores de perfil anotador que apenas completan su juego con acciones distintas a la de anotar puntos. Sin embargo, *Rendimiento* es un índice que pretende no premiar tanto los puntos anotados y darle mayor importancia a otras acciones del juego, muchas de ellas con impacto en el desempeño colectivo. Así, por ejemplo, en la Tabla 3 se tienen los valores de estos dos índices en cuatro casos concretos. En los dos primeros se plantean dos situaciones de dos jugadores que se han limitado a anotar mientras que en las dos segundas la de jugadores que han complementado su juego con una cantidad de acciones más diversa que en los primeros dos casos.

Tabla 3. Cuatro casos concretos de medición mediante Valoración ACB y Rendimiento

| Jugador | Partido | Puntos | Valoración ACB | Rendimiento |
|---|---|---|---|---|
| **Carroll** | Jornada 2 Liga Regular ACB | 25 | 24 | 13.5 |
| **Toolson** | Jornada 24 Liga Regular ACB | 20 | 14 | 6.5 |

---

[3] Para calcular estas medidas se ha ampliado la muestra de 15 jugadores usada hasta ahora y se han tenido en cuenta los 221 jugadores que han disputado al menos 10 partidos durante la liga regular de la ACB en la temporada 2013/14.
[4] Cuando a lo largo del documento se haga referencia a diferencias significativas se ha de tener en cuenta que la correspondiente prueba estadística se ha realizado al 5% de significación.

| | | | | |
|---|---|---|---|---|
| **S. Rodríguez** | Semifinales Super Copa ACB | 14 | 28 | 38.5 |
| **Tomic** | Jornada 1 Liga Regular ACB | 4 | 8 | 16 |

## 3.2 La importancia del tratamiento de los promedios por minuto

Bien es sabido que no es lo mismo obtener una valoración de 20 habiendo jugado 15 minutos o 25. Por tal motivo se opta por promediar las distintas características del juego por minutos, es decir, dividir cada una de ellas por los minutos jugados.

Esta simple operación pone de relieve la producción de cada jugador en función del tiempo jugado. Así por ejemplo, considerando valores de los primeros 15 clasificados en el ranking de Valoración ACB durante la temporada regular 2013/14 podemos observar (ver Tabla 4) que se obtiene que jugadores como Sergio Rodríguez, Nikola Mirotic o Ante Tomic (3 de los 4 jugadores con menos minutos disputados de media) mejoran sustancialmente su posición dentro del ranking de los 15 mejores. En la situación opuesta, bajada de posiciones, se tiene a Andy Panko, Jason Robinson o Tomas Satoransky (3 de los 4 jugadores con más minutos disputados de media). Destacar que Tibor Pleiss ocupa la primera posición en ambos casos.

Tabla 4. Valoración ACB media de los primeros 15 clasificados temporada 2013/14

| **Jugador** | **Equipo** | **Minutos jugados** | **Valoración media** | **Posición** | **Valoración media por minuto** | **Posición** |
|---|---|---|---|---|---|---|
| **Rodríguez** | Real Madrid | 23.01 | 14.55 | 11 | 0.63 | 6 |
| **Doblas** | Gipuzkoa Basket | 29.69 | 13.85 | 15 | 0.45 | 15 |
| **Doelman** | Valencia Basket | 26.74 | 17.33 | 2 | 0.65 | 3 |
| **Gabriel** | Bilbao Basket | 26.01 | 14.76 | 9 | 0.55 | 8 |
| **Kirksay** | Joventut Badalona | 27.48 | 14.42 | 12 | 0.49 | 12 |
| **Llull** | Real Madrid | 28.75 | 15.2 | 7 | 0.52 | 9 |
| **Mirotic** | Real Madrid | 22.62 | 14.7 | 10 | 0.63 | 5 |
| **Mumbrú** | Bilbao Basket | 31.83 | 14.03 | 14 | 0.45 | 14 |
| **Nocioni** | Laboral Kutxa | 25.41 | 14.33 | 13 | 0.56 | 7 |
| **Panko** | Fuenlabrada | 32.46 | 16.91 | 3 | 0.51 | 10 |
| **Pleiss** | Laboral Kutxa | 22.28 | 17.47 | 1 | 0.74 | 1 |
| **Robinson** | Gipuzkoa Basket | 33.81 | 16.21 | 4 | 0.46 | 13 |
| **Rudy** | Real Madrid | 24.82 | 15.64 | 6 | 0.64 | 4 |
| **Satoransky** | Cajasol | 30.75 | 15.81 | 5 | 0.51 | 11 |
| **Tomic** | FC Barcelona | 20.51 | 15.07 | 8 | 0.72 | 2 |

## 4. Medición de la regularidad

Habitualmente se usa el valor medio del índice o característica (puntos, rebotes, etc) a analizar como medida de posición central que resuma los valores observados. Sin embargo, esta medida presenta cierta limitaciones como es la sensibilidad que presenta a los valores extremos. Así por ejemplo, consideremos un Jugador A que anota 5 y 15 puntos y un Jugador B que anota 9 y 11. En ambos casos los dos jugadores tienen una media de 10 puntos. ¿Qué jugador es preferible? En este caso es evidente, el Jugador B. ¿Por qué? Porque es el jugador más regular en su anotación, sus anotaciones están más cerca a su media (por lo que esta es más representativa). Sus datos tienen menos dispersión en definitiva.

Con este ejemplo tan sencillo se pone de manifiesto la necesidad de introducir el concepto de dispersión para poder hablar del jugador o equipo más regular. Si bien Jum & Julius (ver Felipo, 2005) proponen usar el coeficiente de variación (medida estadística para medir la homogeneidad de un conjunto de datos) en este caso proponemos usar su inversa, es decir, la media muestral dividida entre la desviación típica[5] (raíz cuadrada de la varianza). Esto es:

$$\text{Regularidad (Reg)} = \text{Media} * (\text{Desviación Típica})^{-1},$$

de manera que el jugador con mayor regularidad será aquel que presente un mayor valor en este índice (a mayor dispersión menor regularidad).

En el ejemplo anterior, las desviaciones de los dos jugadores son 5 y 1, respectivamente. Luego la regularidad del Jugador A será 2 mientras que la del Jugador B será 10. Atendiendo a este índice será preferible (más regular) el segundo jugador, como ya se había establecido de forma intuitiva.

Luego se trata, por ejemplo, de una medida importante a la hora de determinar qué jugador fichar de entre un grupo de ellos, ya que establece cuál es preferible cuando se parten de características similares. Este último aspecto es de vital importancia: el concepto de regularidad para decidir entre jugadores sólo se ha de usar cuando se parten de valores medios parecidos, ya que en caso contrario se pueden llegar a conclusiones erróneas.

Supongamos que un Jugador A presenta las siguientes valoraciones: 2, 2, 2, 2, 2, 3. Mientras que para un jugador B se tiene: 10, 21, 19, 20, 9, 22. Si se calcula la regularidad en ambos casos se tiene para el primer jugador un valor de 5.307 y para el segundo de 2.914. Luego sería preferible el primero, lo cual no sería la decisión óptima. ¿Qué ha ocurrido? Realmente el primer jugador es más regular que el segundo, sus valores están más próximos al valor medio. Sin embargo, puesto que se parten de valores medios tan distintos (2.166 y 16.833, respectivamente) no se puede usar la regularidad en este caso para decidir cuál de ambos es preferible.

Ya que los rendimientos de los 15 jugadores analizados hasta ahora son muy similares, sería posible usar este concepto para establecer un ranking de regularidad entre ellos. A partir de la Tabla 5 se puede vislumbrar que el jugador más regular es Satoransky. Sin embargo, el mejor jugador de la temporada sería Tomic ya que es el jugador con mayor rendimiento medio por minuto (Satoransky es el séptimo) y el segundo más regular.

---

[5] Adviértase que (normalmente) los paquetes estadísticos ofrecen por defecto la cuasidesviación típica, raíz cuadrada de la cuasivarianza, ya que es el estimador insesgado del verdadero valor (que es desconocido).

También llama la atención el ascenso experimentado por Sergio Rodríguez, inicialmente en el puesto 11 de la Valoración ACB, que pasaría a ser el segundo en cuanto a rendimiento por minuto se refiere y al cuarto más regular.

Tabla 5. Rendimiento por minuto y regularidad de los 15 jugadores mejor valorados en la Liga ACB en la temporada 2013/14

| Jugador | Rendimiento medio por minuto | Desviación típica | Posición | Regularidad | Posición |
|---|---|---|---|---|---|
| Rodríguez | 0.671 | 0.324 | 2 | 2.068 | 4 |
| Doblas | 0.438 | 0.295 | 11 | 1.48 | 11 |
| Doelman | 0.63 | 0.325 | 5 | 1.93 | 5 |
| Gabriel | 0.489 | 0.345 | 10 | 1.41 | 14 |
| Kirksay | 0.49 | 0.28 | 9 | 1.74 | 9 |
| Llull | 0.505 | 0.274 | 8 | 1.83 | 6 |
| Mirotic | 0.596 | 0.333 | 6 | 1.79 | 7 |
| Mumbrú | 0.385 | 0.269 | 14 | 1.43 | 13 |
| Nocioni | 0.394 | 0.309 | 12 | 1.27 | 15 |
| Panko | 0.389 | 0.184 | 13 | 2.11 | 3 |
| Pleiss | 0.657 | 0.37 | 3 | 1.77 | 8 |
| Robinson | 0.369 | 0.251 | 15 | 1.46 | 12 |
| Rudy | 0.644 | 0.402 | 4 | 1.601 | 10 |
| Satoransky | 0.566 | 0.251 | 7 | 2.24 | 1 |
| Tomic | 0.704 | 0.323 | 1 | 2.17 | 2 |

## 6. Importancia de un jugador en el equipo

Hasta el momento se han presentado herramientas para medir el rendimiento y regularidad de un jugador. En teoría, a mejor rendimiento de un jugador mayor ha de ser su influencia en el equipo. Sin embargo, no seríamos capaces de precisar la importancia que tiene un determinado jugador en su equipo.

Se antoja fundamental ser capaces de responder preguntas del tipo: ¿cuál es el rendimiento del jugador en los finales igualados? ¿el mejor o peor comportamiento de un jugador influye en el resultado final de los partidos disputados por su equipo? ¿el rendimiento del jugador varía dependiendo de la competición que disputa? ¿y de si juega como local o visitante? ¿el rendimiento de un jugador es el mismo cuando inicia el partido en el quinteto inicial que cuando sale desde el banquillo? ¿y dependiendo de la entidad del rival?

¿Se pueden responder a estas preguntas sin hacer uso del análisis play by play? La respuesta es afirmativa simplemente tratando la información de forma adecuada.

En la Tabla 6 se tiene el valor medio del +/- (parcial obtenido por un equipo durante la presencia en cancha de un determinado jugador) de los 15 jugadores considerados hasta el momento en todos sus partidos disputados[6], en aquellos con final ajustado (partidos ganados o perdidos por una diferencia inferior o igual a los 5 puntos), en los ganados y perdidos. Por ejemplo, se puede observar que Doblas, Nocioni, Panko, Robinson y Satoransky presentan un promedio de +/- superior a su media en los partidos con final ajustado. También destaca especialmente comprobar que el +/- de Doelman es positivo incluso en aquellos partidos en los que su equipo ha perdido.

Tabla 6. Valor medio del +/- de los 15 jugadores considerados en todos sus partidos disputados, en aquellos con final ajustado, en los ganados y perdidos

| Jugador | +/- total | +/- partidos ajustados | +/- partidos ganados | +/- partidos perdidos |
|---|---|---|---|---|
| **Rodríguez** | 5.93 | -0.88 | 7.13 | -1.67 |
| **Doblas** | 0.15 | 2.5 | 9.63 | -8.28 |
| **Doelman** | 8.43 | 7 | 9.09 | 5.29 |
| **Gabriel** | -0.24 | -1.13 | 11.92 | -6.86 |
| **Kirksay** | -0.88 | -1.89 | 7.69 | -8.94 |
| **Llull** | 7.95 | 1.5 | 9.76 | -3.5 |
| **Mirotic** | 7 | 1.38 | 8.97 | -5.17 |
| **Mumbrú** | -1.41 | -0.25 | 13.5 | -9.55 |
| **Nocioni** | 2.78 | 3.8 | 10.06 | -4.5 |
| **Panko** | -4.79 | 2.11 | 4.58 | -10.14 |
| **Pleiss** | 1.82 | -0.45 | 11.56 | -9.13 |
| **Robinson** | 1.12 | 3.5 | 13.13 | -10.18 |

---

[6] Sería interesante realizar un análisis descriptivo previo de los datos en la búsqueda de datos anómalos que distorsionen los análisis posteriores. De esta forma se obtendría que la medida de posición central (la media) usada para resumir todos los datos sea más representativa.

| | | | | |
|---|---|---|---|---|
| **Rudy** | 9.13 | 1.86 | 13.24 | -13.5 |
| **Satoransky** | -1.46 | 0 | 8.95 | -12.44 |
| **Tomic** | 5.6 | 2.5 | 7.47 | -0.4 |

Ahora bien, toda vez que se deseen realizar comparaciones entre valores medios de dos conjuntos de datos nos debemos de preguntar si existen diferencias significativas desde un punto de vista estadístico. Es decir, ¿existe un patrón repetitivo que expliquen los valores obtenidos o las diferencias se deben simplemente al azar? En tal caso se ha de recurrir a la inferencia estadística, de forma que usando un intervalo de confianza o contraste de hipótesis seamos capaces de decidir si las diferencias son o no significativas.

En la Tabla 7 se tienen los puntos por minuto anotados por cada jugador dependiendo de si su equipo consigue o no la victoria al final del partido. Se puede observar que el comportamiento en cuanto a anotación se refiere es el mismo entre la victoria y derrota excepto para Kirksay, Nocioni y Tomic. En estos tres casos, los jugadores presentan una mejor anotación cuando su equipo gana. Se podría decir por tanto que su producción ofensiva influye en el resultado final del partido.

Tabla 7. Puntos por minuto en función de si el equipo gana o pierde (* indica significativamente distintas a un nivel de significación del 5%)

| **Jugador** | **Puntos por minuto en derrota** | **Puntos por minuto en victoria** |
|---|---|---|
| **Rodríguez** | 0.483 | 0.533 |
| **Doblas** | 0.378 | 0.395 |
| **Doelman** | 0.516 | 0.563 |
| **Gabriel** | 0.511 | 0.465 |
| **Kirksay** | 0.347* | 0.491* |
| **Llull** | 0.492 | 0.476 |
| **Mirotic** | 0.525 | 0.499 |
| **Mumbrú** | 0.448 | 0.532 |
| **Nocioni** | 0.494* | 0.685* |
| **Panko** | 0.543 | 0.568 |
| **Pleiss** | 0.528 | 0.611 |
| **Robinson** | 0.468 | 0.506 |
| **Rudy** | 0.431 | 0.577 |
| **Satoransky** | 0.409 | 0.385 |
| **Tomic** | 0.366* | 0.543* |

Es evidente que las posibilidades que se presentan para comparar los rendimientos de jugadores y analizar su influencia en el desempeño del equipo son numerosas. No es nuestro objetivo exponer de manera exhaustiva todas ellas, sino presentar a la inferencia estadística como una herramienta muy potente para analizar la influencia de un jugador en el desempeño del equipo.

## 7. Análisis empírico

Si bien a lo largo del documento se han usado varios ejemplos con el objetivo de aclarar los distintos conceptos presentados, creemos necesario este apartado de análisis empírico para mostrar las potencialidades de la metodología presentada.

En la Tabla 5 se mostraba el rendimiento por minuto de cada jugador así como su regularidad obteniéndose que el jugador con mejor desempeño sería Tomic (seguido por Sergio Rodríguez) ya que es el que mayor media presenta y el segundo con menos dispersión. Si observamos los valores de los índices defensivos y ofensivos promediados por minutos (ver Tablas 8 y 9) podemos comprobar que en ambos casos este jugador ocupa la segunda posición. En el caso de Rodríguez se ve que si bien es el mejor jugador (primero en promedio y segundo en regularidad) en cuanto al índice ofensivo se refiere, su desempeño defensivo le supone un lastre a la hora de enfrentarse a Tomic.

Una de las ventajas que se presuponían al presentar el rendimiento como la suma de otros dos índices era que se podía medir por separado el desempeño del jugador en su parcela defensiva y ofensiva. En este caso se puede observar que Kirksay es el mejor jugador si hablamos de la defensa. Es más, se podría pensar que su inclusión dentro de los 15 mejores se debe a ella ya que el 46.96% de *Rend* se debe a *ID* (del resto de jugadores no supera el 30% ninguno excepto Robinson con un 37.83%).

Tabla 8. Índice defensivo por minuto medio y regularidad para los 15 jugadores mejor valorados en la liga Regular de la ACB de la temporada 2013/14

| Jugador | ID medio por minuto | Desviación Típica | Posición | Regularidad | Posición |
|---|---|---|---|---|---|
| **Rodríguez** | 0.139 | 0.152 | 8 | 0.917 | 12 |
| **Doblas** | 0.072 | 0.093 | 15 | 0.776 | 15 |
| **Doelman** | 0.187 | 0.129 | 4 | 1.444 | 3 |
| **Gabriel** | 0.106 | 0.117 | 12 | 0.898 | 13 |
| **Kirksay** | 0.231 | 0.155 | 1 | 1.483 | 1 |
| **Llull** | 0.095 | 0.102 | 13 | 0.935 | 10 |
| **Mirotic** | 0.176 | 0.171 | 5 | 1.037 | 7 |
| **Mumbrú** | 0.114 | 0.103 | 10 | 1.106 | 5 |
| **Nocioni** | 0.108 | 0.111 | 11 | 0.98 | 8 |
| **Panko** | 0.072 | 0.088 | 14 | 0.825 | 14 |
| **Pleiss** | 0.187 | 0.197 | 3 | 0.951 | 9 |
| **Robinson** | 0.139 | 0.094 | 7 | 1.476 | 2 |
| **Rudy** | 0.162 | 0.148 | 6 | 1.096 | 6 |
| **Satoransky** | 0.123 | 0.133 | 9 | 0.928 | 11 |
| **Tomic** | 0.193 | 0.155 | 2 | 1.244 | 4 |

Tabla 9. Índice ofensivo por minuto medio y regularidad para los 15 jugadores mejor valorados en la liga Regular de la ACB de la temporada 2013/14

| Jugador | IO medio por minuto | Desviación Típica | Posición | Regularidad | Posición |
|---|---|---|---|---|---|
| **Rodríguez** | 0.531 | 0.301 | 1 | 1.762 | 2 |
| **Doblas** | 0.366 | 0.274 | 10 | 1.332 | 10 |
| **Doelman** | 0.443 | 0.287 | 5 | 1.542 | 6 |
| **Gabriel** | 0.383 | 0.301 | 9 | 1.271 | 11 |
| **Kirksay** | 0.26 | 0.256 | 14 | 1.013 | 13 |
| **Llull** | 0.409 | 0.233 | 8 | 1.755 | 3 |
| **Mirotic** | 0.419 | 0.288 | 7 | 1.456 | 8 |
| **Mumbrú** | 0.271 | 0.236 | 13 | 1.144 | 12 |
| **Nocioni** | 0.286 | 0.288 | 12 | 0.991 | 14 |
| **Panko** | 0.316 | 0.191 | 11 | 1.658 | 4 |
| **Pleiss** | 0.471 | 0.313 | 4 | 1.499 | 7 |
| **Robinson** | 0.229 | 0.245 | 15 | 0.935 | 15 |
| **Rudy** | 0.481 | 0.356 | 3 | 1.351 | 9 |
| **Satoransky** | 0.442 | 0.201 | 6 | 2.162 | 1 |
| **Tomic** | 0.511 | 0.329 | 2 | 1.551 | 5 |

Si optamos por analizar el rendimiento de los jugadores atendiendo al resultado final de partido (ver Tabla 10) obtendremos que Doblas, Mirotic, Mumbrú y Rudy bajan su producción de manera significativa en defensa cuando su equipo pierde, mientras que Rodríguez y Nocioni lo hacen en ataque. De todos estos jugadores, sólo Mumbrú y Nocioni ven disminuido su rendimiento cuando su equipo pierde el partido.

Tabla 10. Rendimiento, índice defensivo y ofensivo medios por minuto de los 15 jugadores mejor valorados en la liga regular de la Liga ACB en función de la victoria o derrota de su equipo (* indica significativamente distintas a un nivel de significación del 5%)

| **Jugador** | **ID en derrota** | **ID en victoria** | **IO en derrota** | **IO en victoria** | **R en derrota** | **R en victoria** |
|---|---|---|---|---|---|---|
| **Rodríguez** | 0.186 | 0.132 | 0.336* | 0.619* | 0.522 | 0.751 |
| **Doblas** | 0.033* | 0.115* | 0.377 | 0.425 | 0.411 | 0.541 |
| **Doelman** | 0.239 | 0.176 | 0.381 | 0.492 | 0.619 | 0.668 |
| **Gabriel** | 0.099 | 0.117 | 0.406 | 0.437 | 0.506 | 0.555 |
| **Kirksay** | 0.249 | 0.209 | 0.211 | 0.374 | .461 | 0.584 |
| **Llull** | 0.091 | 0.096 | 0.527 | 0.417 | 0.617 | 0.514 |
| **Mirotic** | 0.031* | 0.201* | 0.364 | 0.454 | 0.396 | 0.654 |
| **Mumbrú** | 0.083* | 0.172* | 0.281 | 0.381 | 0.365* | 0.553* |
| **Nocioni** | 0.086 | 0.136 | 0.213* | 0.431* | 0.299* | 0.566* |
| **Panko** | 0.063 | 0.087 | 0.343 | 0.339 | 0.407 | 0.427 |
| **Pleiss** | 0.111 | 0.245 | 0.511 | 0.503 | 0.621 | 0.748 |
| **Robinson** | 0.141 | 0.138 | 0.195 | 0.306 | 0.336 | 0.444 |
| **Rudy** | 0.006* | 0.191* | 0.401 | .529 | 0.407 | 0.721 |
| **Satoransky** | 0.085 | 0.159 | 0.485 | 0.482 | 0.571 | 0.642 |
| **Tomic** | 0.146 | 0.207 | 0.465 | 0.571 | 0.612 | 0.778 |

# 7. Conclusiones

El objetivo del presente artículo es medir de forma adecuada el rendimiento y regularidad de un jugador de baloncesto. Por tal motivo, en primer lugar, se presenta un nuevo índice con las siguientes características:

a) Se evita a conciencia el uso del play by play, la pérdida de información se compensa por su accesibilidad para toda persona que desee usarla independientemente de sus recursos (profesionales o aficionados al baloncesto) ya que son fáciles de calcular a partir del boxscore.

b) Se calcula a partir de la suma de otros dos índices que a su vez miden el desempeño del jugador en defensa y ataque. De esta forma se puede detectar cuales son las fortalezas o debilidades en estos dos apartados.

c) Se antoja fundamental promediar los valores obtenidos en función de los minutos jugados.

En segundo lugar, la forma habitual de resumir un conjunto de datos es mediante el uso de la media muestral. Bien es sabido que esta medida de posición central es sensible a valores extremos. Por lo que una exploración previa de los datos en la búsqueda de datos anómalos (partidos ganados o perdidos por gran diferencia, actuaciones puntuales extraordinarias de jugadores, etc) puede ayudar a mejorar la información obtenida. Además, es muy interesante usar el concepto de dispersión (¿cómo de cerca se encuentra los datos a la medida de posición central elegida?). En tal caso, se podría hablar de regularidad del jugador.

Finalmente, en tercer lugar se introduce el uso de la inferencia estadística (intervalos de confianza o contrastes de hipótesis) para establecer comparaciones en determinadas situaciones (clasificadas como éxito o fracaso) de especial interés dentro del deporte del baloncesto. De esta forma se consigue obtener patrones de comportamiento de jugadores o equipos que pueden ser usados con distintos objetivos: scouting de un equipo y jugadores rivales de cara a la preparación de un partido, mejor conocimiento de los puntos fuertes o débiles del propio, información para la toma de decisiones en renovaciones de jugadores o nuevos fichajes, etc. En este caso, las posibilidades que se presentan son numerosas ya que la aplicación de la Estadística al baloncesto no sólo se limita a la inferencia, sino que otras opciones pueden plantearse.

Como futuras líneas de mejora/ampliación del presente trabajo se podría contar, por ejemplo, con:

a) Cálculo de probabilidades condicionadas: ¿cuál es la probabilidad de victoria cuando se gana el rebote? ¿y si se juega a determinado ritmo? ¿y si determinado jugador anota más de cierto número de puntos?

b) Uso del análisis cluster para la clasificación de jugadores a partir de los índices presentados (u otros), de esta forma se obtendrían grupos de jugadores que presenten similitudes entre sí y diferencias con respecto al resto.

c) Uso del concepto de posesión para analizar las decisiones tomadas por cada jugador y, de las mismas, cuáles han conducido a un éxito (es decir, ha sido una decisión correcta).

En el presente trabajo se ha puesto en práctica la metodología expuesta analizando a los 15 jugadores mejor valorados en la liga regular de la Liga ACB durante la temporada 2013/14. Se han observado cambios importantes en los rankings de jugadores dependiendo de la medida que se use, destacando especialmente un mejor posicionamiento en el rendimiento de jugadores como Rodríguez, Satoransky y Tomic o el empeoramiento de Panko. También ha sido posible poner de manifiesto la importancia de la parcela defensiva en la aportación Kirksay, el incremento de la producción ofensiva de este jugador y de Nocioni y Tomic cuando su equipo consigue la victoria o el peor desempeño defensivo de Doblas, Mirotic, Mumbrú y Rudy cuando su equipo pierde.

Los resultados expuestos abre una vía de trabajo muy interesante respecto al impacto individual real del jugador en la competición y los equipos que juega más allá de la común supremacía de la anotación obtenida en el desarrollo de índices de valoración. La revisión del concepto de regularidad puede convertirse en una herramienta muy eficaz a los entrenadores y directivos deportivos de cara a labores de reclutamiento y diseño de los equipos deportivos.

**Bibliografía**